\documentclass[11pt]{article}
\usepackage{hyperref}
\pdfoutput=1
\begin{document}
\title{Gas Bubbles Emerging from a \\Submerged Granular Bed}
\author{John A. Meier\\Joseph S. Jewell\\Christopher E. Brennen
\\\vspace{6pt} Department of Mechanical Engineering \\ California Institute of Technology, Pasadena, CA 91106, USA}
\maketitle

\begin{abstract}
This fluid dynamics video was submitted to the Gallery of Fluid Motion for the 2009 APS Division of Fluid Dynamics Meeting in Minneapolis, Minnesota.  In this video we show some results from a simple experiment where air was injected by a single nozzle at known constant flow rates in the bottom of a granular bed submerged in water.  The injected air propagates through the granular bed in one of two modes.  Mode 1 emergence involves small discrete bubbles taking tortuous paths through the interstitial space of the bed.  Multiple small bubbles can be emitted from the bed in an array of locations at the same time during Mode 1 emergence.  Mode 2 emergence involves large discrete bubbles locally fluidizing the granular bed and exiting the bed approximately above the injection site.  Bead diameter, bead density, and air flow rate were varied to investigate the change in bubble release behavior at the top of the granular bed.\\
This system is a useful model for methane seeps in lakes.  Methane bubbles are released from the decomposition of organic matter in the lake bed.  The initial size of the bubble determines how much of the gas is absorbed into the lake and how much of the gas reaches the surface and is released into the atmosphere.  The size and behavior of the emerging bubbles may also affect the amount of vertical mixing occurring in the lake, as well as the mixing from the lake bed into the benthic layer.
\end{abstract}

The two videos are linked below:\\
\href{http://ecommons.library.cornell.edu/bitstream/1813/14081/3/Meier_APS_GalleryofFluid_mpeg1_final.mpg}{Low Resolution MPEG-1 version}\\
\href{http://ecommons.library.cornell.edu/bitstream/1813/14081/2/Meier_APS_GalleryofFluid_mpeg2.mpg}{High Resolution MPEG-2 version}
\end{document}